\documentclass[prl,twocolumn,amsfonts,superscriptaddress,showpacs]{revtex4}
 
\usepackage[]{graphicx} 
\usepackage{amsbsy} 
\usepackage{float}
\usepackage{color}
\usepackage{subfigure}
\usepackage{amsmath}
 \usepackage{mathtools}

\newcommand{\be}{\begin{equation}}
\newcommand{\ee}{\end{equation}}
\newcommand{\bea}{\begin{eqnarray}}
\newcommand{\eea}{\end{eqnarray}}
\newcommand{\ba}{\begin{eqnarray*}}
\newcommand{\ea}{\end{eqnarray*}}

\begin{document} 
\title{Coupling between a charge density wave and magnetism in an Heusler material}

\author{G.~Lantz}
\affiliation{Institute for Quantum Electronics, Physics Department, ETH Zurich, CH-8093 Zurich, Switzerland}
\author{M.J.~Neugebauer}
\affiliation{Institute for Quantum Electronics, Physics Department, ETH Zurich, CH-8093 Zurich, Switzerland}
\author{M.~Kubli}
\affiliation{Institute for Quantum Electronics, Physics Department, ETH Zurich, CH-8093 Zurich, Switzerland}
\author{M.~Savoini}
\affiliation{Institute for Quantum Electronics, Physics Department, ETH Zurich, CH-8093 Zurich, Switzerland}
\author{E.~Abreu}
\affiliation{Institute for Quantum Electronics, Physics Department, ETH Zurich, CH-8093 Zurich, Switzerland}
\author{K.~Tasca}
\affiliation{Institute for Quantum Electronics, Physics Department, ETH Zurich, CH-8093 Zurich, Switzerland}
\author{C.~Dornes}
\affiliation{Institute for Quantum Electronics, Physics Department, ETH Zurich, CH-8093 Zurich, Switzerland}
\author{V.~Esposito}
\affiliation{Swiss Light Source, Paul Scherrer Institut, CH-5232 Villigen PSI, Switzerland}
\author{J.~Rittmann}
\affiliation{Swiss Light Source, Paul Scherrer Institut, CH-5232 Villigen PSI, Switzerland}
\author{Y.W.~Windsor}
\affiliation{Swiss Light Source, Paul Scherrer Institut, CH-5232 Villigen PSI, Switzerland}
\author{P.~Beaud}
\affiliation{Swiss Light Source, Paul Scherrer Institut, CH-5232 Villigen PSI, Switzerland}
\author{G.~Ingold}
\affiliation{Swiss Light Source, Paul Scherrer Institut, CH-5232 Villigen PSI, Switzerland}
\author{S.L.~Johnson}
\affiliation{Institute for Quantum Electronics, Physics Department, ETH Zurich, CH-8093 Zurich, Switzerland}
\date{\today} 
 
\begin{abstract} 
The Prototypical magnetic memory shape alloy Ni$_2$MnGa undergoes various phase transitions as a function of temperature, pressure, and doping. In the low-temperature phases below 260 K, an incommensurate structural modulation occurs along the [110] direction which is thought to arise from softening of a phonon mode. It is not at present clear how this phenomenon is related, if at all, to the magnetic memory effect. Here we report time-resolved measurements which track both the structural and magnetic components of the phase transition from the modulated cubic phase as it is brought into the high-symmetry phase. The results suggest that the photoinduced demagnetization modifies the Fermi surface in regions that couple strongly to the periodicity of the structural modulation through the nesting vector. The amplitude of the periodic lattice distortion, however, appears to be less affected by the demagnetizaton.
\end{abstract} 
 
\maketitle 

The magnetic shape memory effect (MSME) in Heusler materials is characterized by a very large magnetic field-induced strain that can exceed 11\% \cite{Sozinov2002,Pagounis2015}. The particularly large MSME seen in Ni-Mn-Ga alloys is promising for technological applications \cite{Gabdullin2016,Pagounis2015,Sozinov2013}. Stoichiometric Ni$_2$MnGa exhibits MSME only in its lowest temperature phase. At higher temperatures Ni$_2$MnGa undergoes a series of structural and magnetic phase transitions. The low-temperature phase, characterized by a  martensitic tetragonal structure (MT) with an incommensurate structural modulation, is stable up to T$_\text{MT}$ = 220 K. Between 220 K and 260 K lies the so-called pre-martensitic (PMT) phase which can be described as a cubic austenitic phase with an incommensurate structural modulation. Between 260 K and 380 K this modulation disappears and the material is in the austenitic cubic phase (AUS).  All the phases below the Curie temperature of 380 K are ferromagnetic.

The MSME is widely understood to originate from the reorientation of twinned domains in the tetragonal phase. The microscopic mechanism of the structural modulation and its link to both the MSME and ferromagnetism are, however, not completely understood. Since the modulation, upon cooling, arises already in the PMT phase, understanding its origin in this phase might shed light on its potential link to the MSME. Studying the modulation in the PMT phase presents the advantage of maintaining the same lattice parameters during the PMT-AUS phase transition, which allows sensitivity to only the modulation dynamics. The modulation is incommensurate and is thought to be triggered by a phonon softening along the [110] direction \cite{Zheludev1995,Mariager2014a}. The modulation can be understood within two scenarios. The first is an adaptive model where the modulation is considered as an effective twinning arising from minimization of elastic energy \cite{Khachaturyan1991,Kaufmann2010}. The other scenario is a Fermi surface nesting which results in a phonon softening and the appearance of a charge density wave (CDW). The latter has recently gotten more experimental and theoretical support \cite{Bungaro2003,Shapiro2007,DSouza2012,Singh2015,Belesi2015}. A unified theory has been proposed by Gruner et al., in which the adaptive nanotwinning is locked on the CDW \cite{Gruner}.
Schubert et al. \cite{Schubert2015} have shown that doping with magnetic cobalt impurities causes a change in both the spatial periodicity of the modulation and the associated amplitude mode associated with the modulation. While this is indicative of a possible link between magnetism and the CDW,  Ni-Mn-Ga alloys are in general very sensitive to chemical doping and the changes observed could also be caused by local structural or electronic modifications \cite{Planes2009}.

Time-resolved “pump-probe” experiments offer the opportunity to study the coupling of the magnetic and structure in materials that have been subjected to a sudden perturbation. Previous pump-probe experiments on Ni$_2$MnGa have been able to track the amplitude mode of the CDW across the MT-AUS phase transition \cite{Mariager2012,Mariager2014,Schubert2015}.  
In this letter, we study the dynamics of stoichiometric Ni$_2$MnGa starting from the PMT phase using time-resolved magnetic optical Kerr effect (MOKE) and time-resolved X-ray diffraction (trXRD). This combination of techniques enables us to follow separately the structural modulation and the magnetism after strong electronic excitation from a femtosecond pulse of light. 

The same sample was used for all our measurements, a single crystal of stoichiometric Ni$_2$MnGa  from AdaptaMat Ltd. oriented with the [001] direction normal to the surface and has a 5M structure in the MT phase. The PMT and MT transition temperatures were measured at T$_\text{PMT} = 264$~K and T$_\text{MT} = 224$~K, as reported in Ref. \cite{Mariager2012}.

The MOKE experiment was performed in a polar configuration with nearly collinear pump-probe along the surface normal in a 0.7 T field, which is enough to saturate the magnetization (see Supplemental Material). A 250 kHz Ti-sapphire laser system was used to deliver 45 fs long 800 nm pump pulses and 400 nm frequency doubled probe pulses. The probe and pump attenuation lengths were 16 nm and  23 nm, respectively. The pump-probe traces were measured with heterodyne detection and with both magnetization directions. In both experiments, the sample temperature was set to 225 K, well within the PMT phase, using a nitrogen cryogenic blower. 

The trXRD experiment was performed at the FEMTO slicing source of the Swiss Light Source (PSI, Villigen, Switzerland)\cite{Beaud2007}. The X-ray beam was in grazing incidence at an angle of 0.65$^\circ$ and at an energy of 7 keV, resulting in an attenuation length of 55 nm. The 800 nm pump pulse had an incidence angle of 12$^\circ$ resulting in an attenuation length depth of 21 nm \cite{Lee2002}. The penetration depth mismatch between pump and probe leads to a smearing of the observed dynamics that has to be accounted for in the modeling. The overall time resolution for these measurements was better than 150 fs \cite{Beaud2007} (see Supplemental Material). The sample was previously oriented at the Material Science beamline of the Swiss Light Source \cite{Willmott2013}, where the orientation matrix was computed and then verified during the time-resolved experiment. An avalanche photodiode was used to measure the X-ray intensity for each pulse.

\begin{figure}
\includegraphics[angle=0,width=1\linewidth,clip=true]{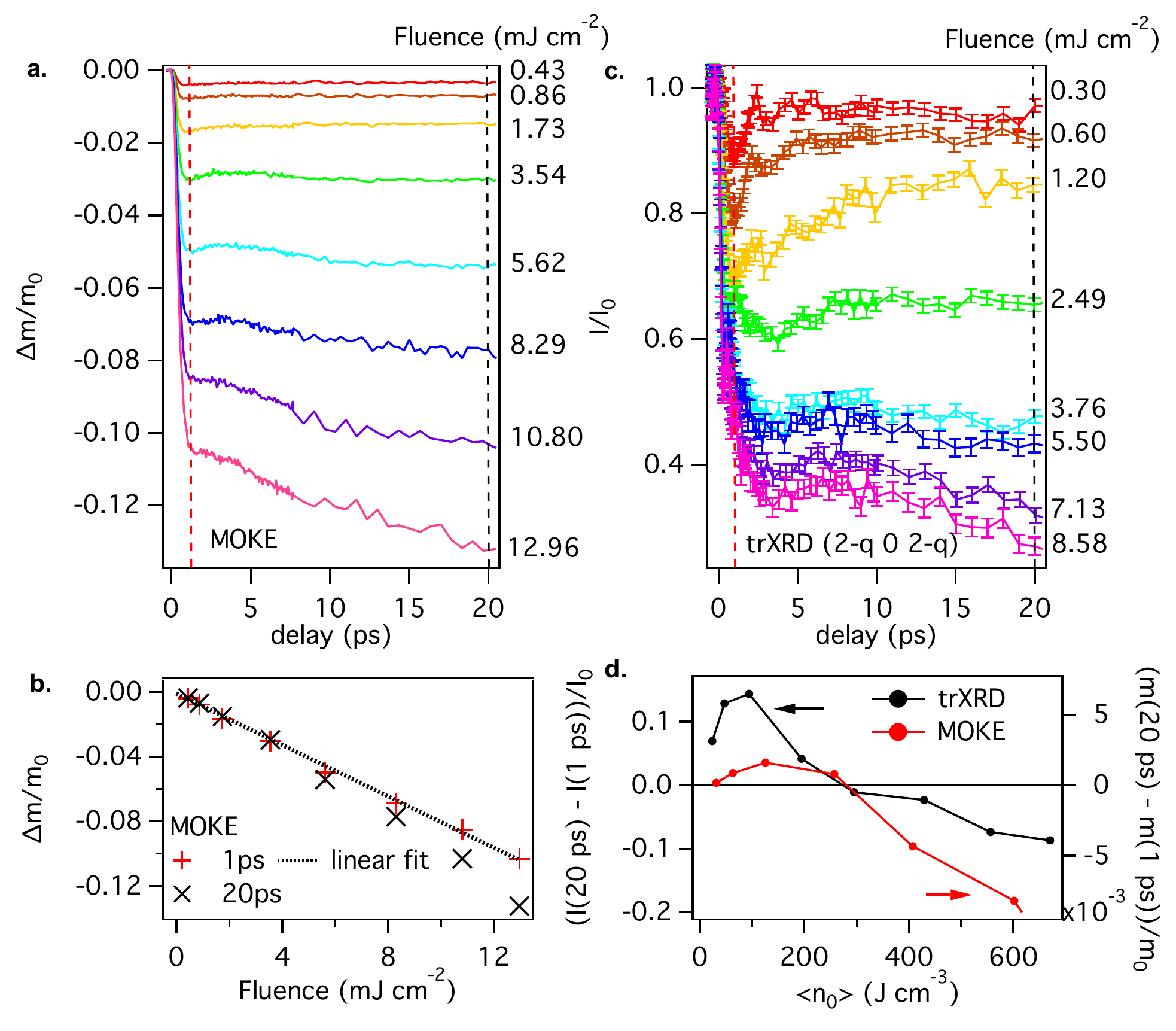}
\caption{Comparison between MOKE and trXRD. \textbf{a.} Relative changes in the MOKE signal for difference fluences. The red and black dotted lines show the 1 ps and 20 ps delays. \textbf{b.} Difference in magnetization at 1 ps and 20 ps versus fluence. The dotted line shows the linear behavior versus fluence of the demagnetization at 1 ps delay and the nonlinearity of the 20 ps delay. \textbf{c.} (2-q 0 2-q)  intensity for different fluences. \textbf{d.} Difference in demagnetization and loss of intensity for MOKE and trXRD between 20 ps and 1 ps delays versus observed excitation density $\left< n_0 \right>$. The difference is positive up to 300 J cm$^{-3}$ and becomes negative after for both MOKE and trXRD.} 
\label{trmoke} 
\end{figure}

Fig. \ref{trmoke}a shows an overview of the magnetization dynamics for several fluences. The magnetization initially drops over a timescale of 1 ps. This is followed by a second timescale of about 10 ps over which the magnetization either partially recovers by a small amount (at low fluences) or further decreases (at high fluences). The fluence dependence of the relative change in magnetization is shown in Fig.~\ref{trmoke}c at both 1~ps and 20~ps after excitation. The dotted line is a linear fit for demagnetization versus fluence. The changes at 1~ps agree well with this model, whereas the later time changes do not.

For comparison, Fig.~\ref{trmoke}c shows the time-dependent changes in the X-ray intensity at the maximum of the rocking curve of the (2-q 0 2-q) superlattice reflection, where q = 0.341 \cite{Mariager2014a}, over a similar range of excitation fluences as in Fig.~\ref{trmoke}a. The dynamics of the superlattice reflection are more complex, but we can also see a fast decrease in intensity over the first few picoseconds followed by slower timescale dynamics that are somewhat reminiscent of the behavior of the magnetization.  Because the penetration depth mismatch between the probe and the pump, the trXRD measurement is an average over regions of the crystal with very different electronic excitation densities. In order to quantitatively compare the excitation density dependence of the MOKE and trXRD measurements, we convert the incident fluence to the average excitation density $\left< n_0 \right>$ seen by the probe (see Supplemental Material). Fig.~\ref{trmoke}d shows the normalized difference in measured intensities at 20 ps and at 1 ps for both MOKE and trXRD measurements, as a function of $\left< n_0 \right>$.  We see similar behavior:  at excitation densities below $300$~J cm$^{-3}$ both the magnetization and diffracted intensity partially recover, whereas above this value they both further decrease over this time scale.
  
\begin{figure}
\includegraphics[angle=0,width=1\linewidth,clip=true]{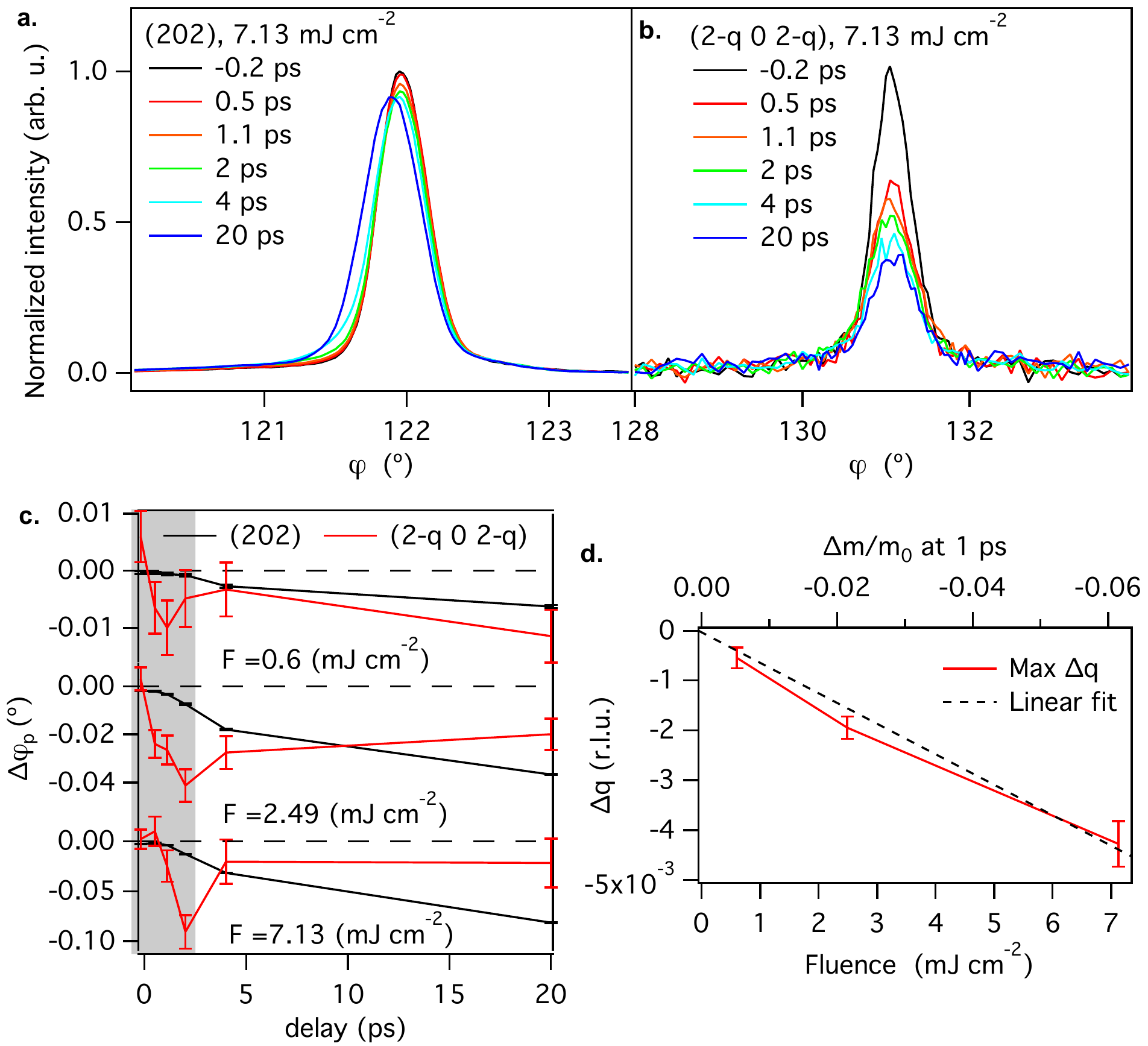}
\caption{Rotation scans at a 7.13 mJ cm$^{-2}$ fluence for the \textbf{a.} (202) Bragg reflection and \textbf{c.} (2-q 0 2-q) superlattice reflection. \textbf{b.} Peaks positions versus time for different fluences. The (202) does not shift significantly before 4 ps whereas the (2-q 0 2-q) reaches its minimum around 2 ps. The gray region represents the times for which the orientation matrix is assumed to be nearly unchanged in order to estimate $\Delta q$ of the superlattice distortion \textbf{d.} Extracted maximum $\Delta q$ change versus fluence (bottom axis) and demagnetization value at 1 ps from Fig. \ref{trmoke}c (top axis).} 
\label{rot} 
\end{figure}

In order to study possible time-dependent changes in the wavevector of the modulation, we show in Fig.~\ref{rot} the diffracted intensity as a function of rotation angle $\varphi$ about the surface normal for different time delays for both the (202) and its satellite reflection (2-q 0 2-q). A fit using a Voigt function was performed to extract the precise peak position, $\varphi_p$. Fig. \ref{rot} c. shows the resulting $\varphi_p$ values. We first consider the (202) Bragg reflection. $\varphi_p$ of the (202) reflection decreases monotonically, consistent with an uniaxial expansion of the crystal along the direction of the surface normal that is typical for laser-driven heating~\cite{Johnson2010}. This expansion takes place over a timescale determined by the strain wave propagation from the surface of the crystal through the probed region, that is roughly $\delta/v=  12.5$~ps, with probe depth $\delta = 55$~nm and the speed of sound $v = 4.4$ nm ps$^{-1}$ \cite{Stenger1998}. Over the first two picoseconds the value of $\varphi_p$ shifts less than 0.02$^\circ$ even at the highest fluence.  At 20~ps delay the strain can be estimated at 0.6$\%$ for the maximum fluence of 7.13 mJ cm$^{-2}$~\cite{Johnson2010}. The timescale of the expansion matches the tens of picoseconds timescale seen in the magnetization dynamics from the MOKE measurements and in the change in the overall intensity of the satellite reflection (Fig. \ref{trmoke}). At high fluences this appears to be consistent with equilibrium measurements that show that both the modulation and the magnetization decrease with increasing strain along the [001] direction \cite{Nie2015,Planes1997}, although at low fluences we see these are anticorrelated in the time-resolved measurement which makes it difficult to draw a definite conclusion about the relationship between strain, magnetization and the modulation amplitude in the nonequilibrium case. 

In marked contrast to the dynamics of $\varphi_p$ for the (202) reflection, $\varphi_p$ for the satellite reflection shows changes that are much larger and are maximal  at 1-2 ps after excitation, after which $\varphi_p$ relaxes back toward its initial unexcited value.  This suggests that the value of $q$ undergoes changes that are independent of the dynamics of the underlying unit cell dimensions. Under the rough assumption that over the first few picoseconds these unit cell dimensions remain at their initial, unexcited values, and that the direction of the modulation vector is unchanged, we can relate the measured changes in $\varphi_p$ to changes in the magnitude of the superlattice modulation wavevector $q$. Fig.~\ref{rot}d shows $\Delta q$, the maximum value of the time-dependent change in $q$, as a function of fluence. The value of $\Delta q$ is approximately proportional to the fluence, as can be seen by the dashed linear fit to the data.  Note also that $\Delta q $ decreases with increasing fluence, a trend which is opposite to that seen on increasing the temperature~\cite{Singh2015}. The measured $\varphi_p(t)$ is qualitatively different from the time-dependence of the intensity, which gives us some confidence that the apparent small changes are indeed reflective of a genuine change in the modulation wavevector.

Both the rapid onset of these changes in $q$ as well as their linearity with respect to fluence are quite similar to the initial behavior of the magnetization dynamics as seen in the MOKE measurements. The magnetization drops on a time scale of 1 ps, which is slightly faster than the observed maximum deviation in $q$ seen around 2 ps for most fluences. On this basis we suggest that the demagnetization drives the $q$ shift. In equilibrium, Lee et al.~\cite{Lee2002a} have shown that changes in the magnetization density lead to proportional changes in q. In Fig.~\ref{rot}d, we show on the top axis the experimental demagnetization value at 1 ps from Fig. \ref{trmoke}b for an equivalent fluence. Using the theoretical $q$ shift versus magnetization from Ref. \cite{Lee2002a} and assuming that the sample is initially at 80\% magnetization saturation, we calculate that a fluence of 7.13 mJ cm$^{-2}$ leads to a $\Delta q$ of -0.016, which is 4 times lower than the measured trXRD value. Despite this quantitative disagreement, the signs of shifts and their linearity are consistent with the idea of the demagnetization playing some role. The quantitative discrepancy may be related to additional contributions to the Fermi surface nesting arising from the highly excited electronic states. The decrease of q with demagnetization closely resembles the increase of the real space periodicity in the phase transitions from 5M to 7M to non modulated structures that occur during compression stress \cite{Pagounis2015}.

\begin{figure}
\includegraphics[angle=0,width=1\linewidth,clip=true]{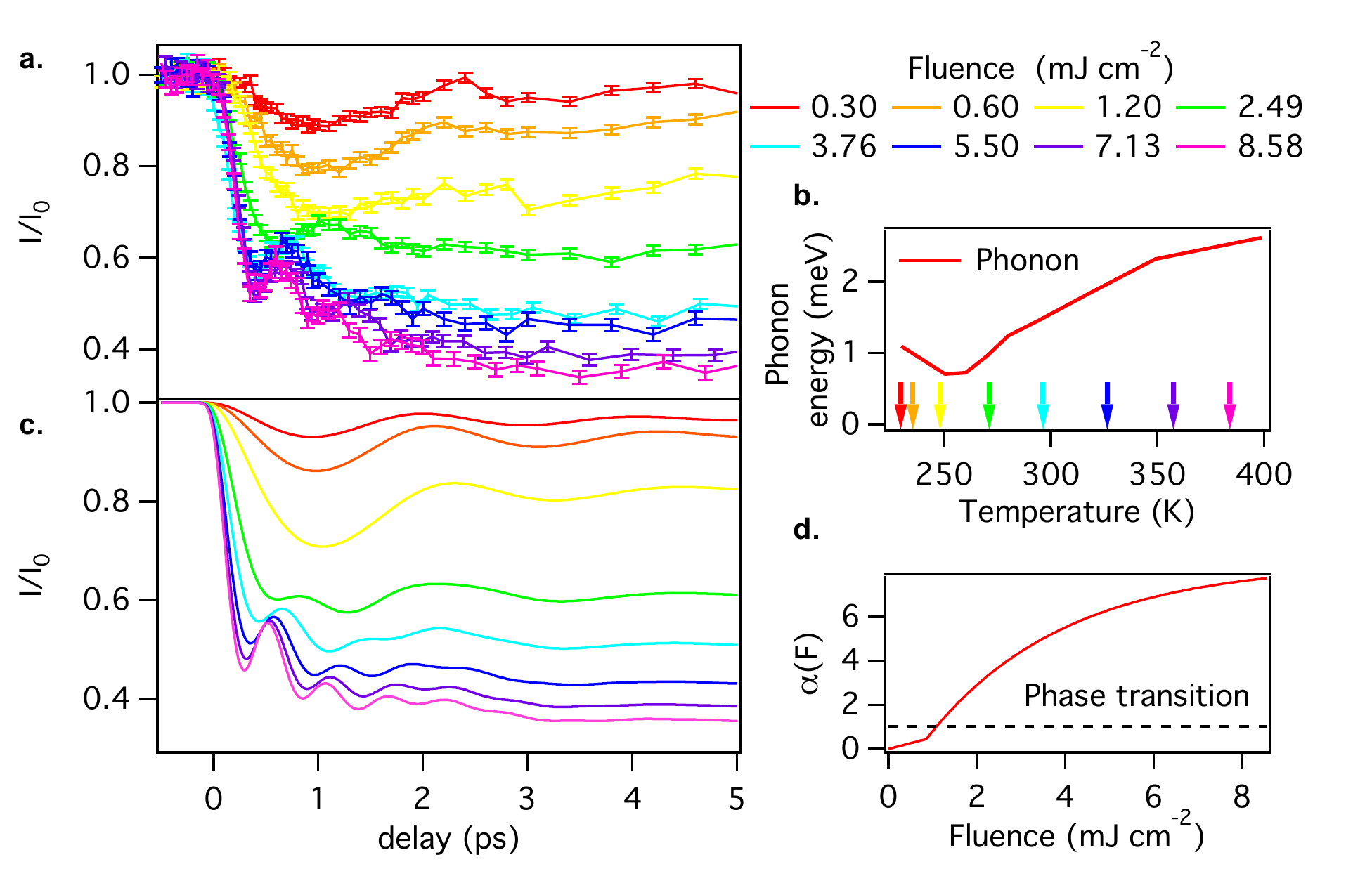}
\caption{trXRD intensity compared to single parameter model. \textbf{a.} trXRD time scan of the superlattice reflection for several fluences. \textbf{b.} Phonon energy versus temperature, showing the softening (adapted from \cite{Zheludev1995}). The colored arrows show the average estimated final temperature reached by the excited area corresponding to different fluences. \textbf{c.} Fitted single parameter model. \textbf{d.} Fluence dependence of the excitation function, $\alpha(\text{F})$. } 
\label{model} 
\end{figure}

We now turn our attention to the ten first picoseconds of the satellite reflection intensity, which tracks the magnitude of the structural modulation, summarized in Fig.~\ref{model}a, which is a zoomed in version of Fig.~\ref{trmoke}c. First, we note that the time-dependent changes are markedly nonlinear with respect to the absorbed fluence. At fluences of 1.2 mJ cm$^{-2}$ and below, the dynamics appear largely consistent with a model where the intensity changes can be described as a displaced coherent excitation~\cite{Zeiger1992} of the amplitude mode of the CDW, where the displaced mode has an approximate frequency $\nu = 0.35$~THz. As the fluence increases beyond 2~mJ cm$^{-2}$, however, the time-dependence changes dramatically. Most notably the frequency of oscillation increases by up to a factor of 4 relative to the low-fluence behavior. This increase in frequency with fluence stands in contrast to the temperature dependence of the amplitude mode frequency, where an increase in temperature causes a decrease in frequency \citep{Schubert2015}. A qualitatively similar increase in frequency in the coherent oscillations of a superlattice reflection has been previously seen in other structurally modulated materials~\cite{Huber2014,Beaud2014}; in these two cases a doubling of the frequency is reported and was attributed to an excitation-induced change in the symmetry of the interatomic potential energy surface consistent with a phase transition to a higher symmetry state. In our case we see a much more dramatic increase in frequency, but we can model this behavior in a similar way.

We consider that locally the magnitude of the structural modulation can be represented by a single variable $\eta$. To treat the dynamics of the phase transition we consider an effective potential energy which has the same functional dependence on $\eta$ as does the Landau free energy. This should in principle correspond to the phase transition order, which is first order for the PMT to AUS transition in equilibrium conditions. However there is no measurable hysteresis and the latent heat of the transition is very small. Therefore we can consider that the transition is close to second order \cite{Planes1997,Velikokhatnyi1999}. We assume that the contribution of $\eta$ to the potential energy is of the form:
\begin{equation}
\label{free_e}
\begin{aligned}
V(\eta)=\frac{1}{2}(\alpha(\text{F})-1)\eta^2+\frac{1}{4}\eta^4,
\end{aligned}
\end{equation} 
where $\alpha(\text{F})$ is a function of the fluence F. The shape of the potential has its minimum at $\eta=1$ when $\alpha(0)=0$ in order to use a normalized intensity. The time-dependence is computed using the following equation of motion: 
\begin{equation}
\label{osc}
\begin{aligned}
\frac{1}{\omega_{DW}^2}\frac{\partial^2}{\partial^2 t}\eta+\frac{\partial}{\partial \eta}V(\eta)
+\frac{\gamma}{\omega_{DW}^2}\frac{\partial}{\partial t}\eta=0,
\end{aligned}
\end{equation}
where $\omega_{DW}$ is the frequency in the double well potential at $\alpha=0$ and $\gamma$ is the damping rate of the motion. The diffracted X-ray intensity I is then proportional to $\eta^2$. Initially $\alpha(0)=0$, which is changed instantaneously to $\alpha(\text{F})$ at $t_0$, the arrival time of the pump pulse. This sudden change causes the system to oscillate in the new potential, which is fully determined by the function $\alpha(\text{F})$.

In Refs \cite{Huber2014} and \cite{Beaud2014} the excitation function, $\alpha$ is assumed to be proportional to F, in close analogy to the temperature dependence of the Landau theory \cite{Planes1997,Buchelnikov2000,Gooding1989}. Some modifications of this model are needed to account for the stronger increase in frequency observed in the present case. As a guide to the functional form of $\alpha(\text{F})$, we consider the experimentally measured temperature dependence of the phonon frequency associated with the amplitude mode, shown in Fig. \ref{model}b. Here we have marked the temperatures which correspond to the various excitation fluences used, under the assumption that the energy density deposited by the laser fully equilibrates with the lattice and that transport effects are negligible \cite{Uijttewaal2009} (Supplemental Material). The three temperatures corresponding to the three lowest fluences are below the phase transition which could explain the sudden change of the observed dynamics above 2 mJ~cm$^{-2}$. We therefore use the phonon energy versus temperature to guide the construction of $\alpha(\text{F})$. The function is as defined below, linear up to a critical fluence and exponential above:
\begin{equation}
\begin{aligned}  
                \begin{array}{ll}
                  \alpha(\text{F})=a\text{F} \quad\text{for}\quad  \text{F}<\text{F}_\text{c}\\
                  \alpha(\text{F})=a \text{F}_\text{c} + b(1-e^{-(\text{F}-\text{F}_\text{c})/d})\quad\text{for}\quad \text{F}>\text{F}_\text{c}\\
                \end{array}
\end{aligned}
\end{equation}
where $a$, $b$, $d$, and $\text{F}_\text{c}$ are constants. We perform a global fit of all curves simultaneously, using $a$, $b$, $d$, $\text{F}_\text{c}$, $\omega_{DW}$, and $\gamma$ as fitting parameters. We fit the first 10 ps and ignore the strain contribution. Since the intensity of the laser decays exponentially inside the material, the absorbed energy has a strong depth dependence. In order to account for this penetration depth we divide the material into layers of thickness $\delta$ = 1 nm and calculate the absorbed energy for each of them. Each layer has a new potential, determined by $\alpha(\text{F})$, which is then used in the equation of motion. Since the correlation length of the material is small, different layers contribute incoherently to the signal such that the global intensity is the sum over all the layers by weighing them as a function the X-ray penetration depth.

The best fit result is shown in Fig. \ref{model}c using $\alpha(\text{F})$ shown in Fig. \ref{model}d. The dynamics are well reproduced. It is worth noticing that $\alpha(\text{F}_\text{c})<1$, which is lower than the excitation parameter that triggers the phase transition. Similarly, the soft phonon at equilibrium does not reach zero at $T_\text{PMT}$ but by interpolation would reach zero at $T_\text{PMT}$ - 5 K. The excitation parameter $\alpha(\text{F})$ is nonlinear and follows a similar behavior as the phonon softening, showing that the amplitude of the modulation is mostly uncorrelated to the magnetism, which is linear versus fluence and is not needed to model the amplitude dynamics.

In conclusion, we have probed the magnetization and the structural modulation separately using MOKE and trXRD starting from the PMT phase respectively, allowing us to analyze the correlation between the two. We have shown that strain propagation contributes to shifting the Bragg reflection and lowering the magnetization on a timescale of tens of picoseconds. On a shorter timescale the superlattice peak shifts linearly with fluence. We suggest that the demagnetization happening after photoexcitation modifies the Fermi surface, which in turn changes the magnitude of the nesting vector and therefore the modulation periodicity. In contrast, amplitude of the modulation has a very nonlinear fluence behavior that can be modeled with a single variable, $\eta$, potential energy fully determined by an excitation parameter, which has a similar trend in fluence than the phonon softening has in temperature. It thus appears that the magnetization has some impact on the periodicity of the lattice distortion but does not strongly determine the amplitude of the distortion. The possible correlation between the magnetism and the modulation periodicity may be crucial for understanding the precise mechanism of the MSME.

{\em Acknowledgments.} Time resolved x-ray diffraction measurements were carried out at the X05LA beam line, and preparative static grazing incidence diffraction measurements were performed at the X04SA beam line of the Swiss Light Source, Paul Scherrer Institut, Villigen. We acknowledge financial support by the NCCR Molecular Ultrafast Science and Technology (NCCR MUST), a research instrument of the Swiss National Science Foundation (SNSF). G. L. also acknowledges the financial support of ETH Career Seed Grant SEED-80 16-1. E. A. acknowledges support from the ETH Zurich Postdoctoral Fellowship Program and from the Marie Curie Actions for People COFUND Program.

\bibliographystyle{apsrev}

\subsection{Details on the time-resolved X-ray diffraction experiment}
The femtosecond X-ray experiment is performed on the FEMTO slicing source of the Swiss Light Source (PSI, Villigen, Switzerland)\cite{Beaud2007}. The femtosecond X-ray pulse used is superimposed on a $\sim50$~ps background pulse, which contributes to about 21$\%$ of the intensity. This picosecond background is systematically recorded looking at X-ray pulse from the single electron bunch arriving just before the sliced bunch (1 $\mu$s before). For the scans as a function of rotation angle $\varphi$,  we also perform pump-probe scans using as a probe the x-ray pulse after a time when the femtosecond portion of has relaxed, leaving only the long-time background pulse. These scans are subtracted from the rotation scans in order to avoid any artifact arising from the picosecond background.

The sample was previously oriented at the Material Science beamline of the Swiss Light Source \cite{Willmott2013}, where the orientation matrix was computed using the Bragg reflections (202), (-202), (0-22), and the (2-42). During the time-resolved experiment the orientation matrix was verified using the reflections (202) and (2-q 0 2-q), where q = 0.341. 

\subsection{Excitation density calculations}
The X-ray, the optical 800 nm pump, and the 400 nm probe do not have the same penetration depth, therefore instead of using the absorbed fluence we use the average excitation density. We divide the sample into n layers of thickness $dz$ starting at the surface with i ranging from 0 to n. The excitation density for the i$^{th}$ layer is:
\begin{equation}
n(i)=\frac{F(1-R)(e^{\frac{-i dz}{\alpha_L}}-e^{\frac{-(i+1) dz}{\alpha_L}})}{dz}
\end{equation}
where F is the incoming fluence, R is the reflectivity of the pump laser at the specific angle of the experiment (R = 0.06 for trXRD and R = 0.11 for MOKE \cite{Lee2002}), $\alpha_L$ is the penetration depth of the laser (23 nm for MOKE and 21 nm for trXRD). Here $F(1-R)$ is the absorbed fluence.\\
We then compute the average excitation density seen by the probe:
\begin{equation}
\left< n_0 \right>=\sum_{i=0}^{n}n(i)\frac{1}{\alpha_p} e^{-idz/\alpha_p}
\end{equation}
where $\alpha_p$ is the penetration depth of the probe (55 nm for trXRD and 16 nm  for MOKE).
\subsection{Static magnetic measurement}
\begin{figure}
\includegraphics[angle=0,width=1\linewidth,clip=true]{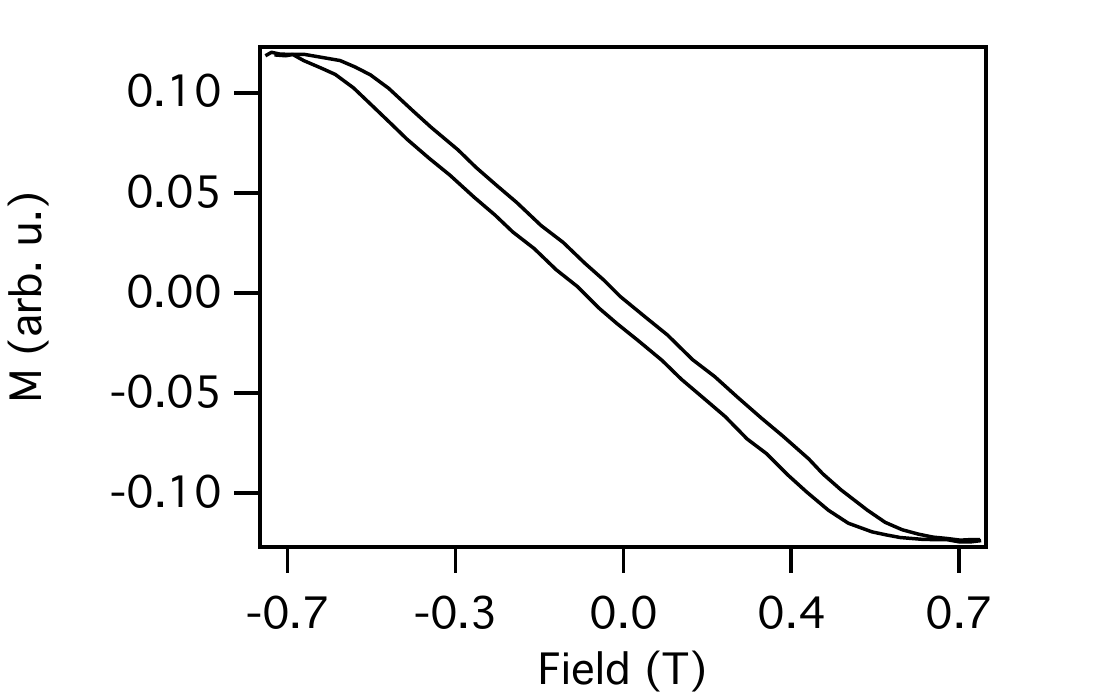}
\caption{Magnetization of bulk sample using static magnetic optical Kerr effect.} 
\label{hyst} 
\end{figure}

In order to verify that the sample magnetization is saturated for the pump-probe experiment, we perform static magnetic optical Kerr effect with no pump pulse. Fig. \ref{hyst} shows that the field of 0.7 T is enough to saturate completely the bulk sample. Therefore we have performed the pump-probe experiment using 0.7 T.
\subsection{Temperature calculations}
Neglecting transport effects, we estimate the temperature after achieving local thermal equilibrium as :
\begin{equation}
T_f=T_i+\frac{\left< n_0 \right>}{C_p}
\end{equation}
Where $T_f$ and $T_i$ are the final and initial temperature, and $C_P$ is the sample's heat capacity (95 J K$^{-1}$ mol$^{-1}$ \cite{Uijttewaal2009}). The estimated final temperatures are mentioned in the main text. We neglect heat diffusion since it has a usual timescale of hundreds of picoseconds to nanoseconds, which is outside our measurement window.

\end{document}